\documentclass[twocolumn,english]{revtex4-1}
\usepackage[T1]{fontenc}
\usepackage{color}
\usepackage{amsmath}
\usepackage{amssymb}
\usepackage{graphicx}
\usepackage{esint}
\usepackage{lineno}

\makeatletter
\@ifundefined{textcolor}{}
{%
 \definecolor{BLACK}{gray}{0}
 \definecolor{WHITE}{gray}{1}
 \definecolor{RED}{rgb}{1,0,0}
 \definecolor{GREEN}{rgb}{0,1,0}
 \definecolor{BLUE}{rgb}{0,0,1}
 \definecolor{CYAN}{cmyk}{1,0,0,0}
 \definecolor{MAGENTA}{cmyk}{0,1,0,0}
 \definecolor{YELLOW}{cmyk}{0,0,1,0}
}

\renewcommand{\fnum@figure}{\textbf{Figure~\thefigure}}
\usepackage{units}
\usepackage{lipsum}

\usepackage{babel}

\makeatother

\usepackage{babel}
\begin{document}

\title{Exciton Relaxation Cascade in Two-dimensional Transition-metal dichalcogenides}

\author{Samuel Brem$^1$}
\email{samuel.brem@chalmers.se}
\author{Gunnar Berghaeuser$^1$}
\author{Malte Selig$^2$}
\author{Ermin Malic$^1$}
\affiliation{$^1$Chalmers University of Technology, Department of Physics, 41296 Gothenburg, Sweden}
\affiliation{$^2$Technical University Berlin, Institute of Theoretical Physics, 10623 Berlin, Germany}

\begin{abstract}
Monolayers of transition-metal dichalcogenides (TMDs) are characterized by an extraordinarily
strong Coulomb interaction giving rise to tightly bound excitons with binding energies of hundreds of meV. Excitons dominate the optical response as well as the ultrafast dynamics in TMDs. As a result, a microscopic understanding of exciton dynamics is the key for technological application of these materials.  
In spite of this immense importance, elementary processes guiding the formation and relaxation of excitons after optical excitation of an electron-hole plasma has remained unexplored to a large extent. 
Here, we provide a fully quantum mechanical description of momentum- and energy-resolved exciton dynamics in monolayer molybdenum diselenide (MoSe$_2$) including optical excitation, formation of excitons, radiative recombination as well as phonon-induced cascade-like relaxation down to the excitonic ground state.
Based on the gained insights, we reveal experimentally measurable features in pump-probe spectra providing evidence for the exciton relaxation cascade.

\end{abstract}
\maketitle

\section{Introduction}
Atomically thin semiconductors, such as transition metal dichalcogenides (TMDs), have revolutionized research in optics and electronics \cite{mak2016photonics,schaibley2016valleytronics,geim2013van,lopez2013ultrasensitive, pospischil2016optoelectronic}. Alongside with the tremendous potential for innovative technologies, their optical and electronic properties further allow to study fundamental many-particle processes in an unprecedented scope. In particular, the reduced dimensionality of these materials leads to a weak dielectric screening, resulting in an optical response dominated by the emergence of excitons, i.e. quasi-particles comprised of Coulomb-bound electron-hole pairs \cite{Mak2010,ugeda2014giant,Li2014,steinhoff2015efficient, wang2017excitons}. Equivalent to hydrogen orbitals, excitons exhibit discrete internal degrees of freedom \cite{hill2015observation,chernikov2014exciton}, which makes them appear as a semiconductor analogon of atoms in a gas, but with properties which can be easily externally tuned \cite{raja2017coulomb, conley2013bandgap, steinhoff2014influence} by e.g. changing the dielectric environment. The microscopic understanding and controlled manipulation of the internal structure of excitons could enable an entire new class of light emitters and absorbers. 

Although, the analogy to atoms might help to illustrate some exciton properties, the formation and relaxation of these collective excitations are exclusively guided by many-particle effects \cite{koch2006semiconductor, Kira2006, Thranhardt2000}. Despite the continuous progress made in 2D material research during the last years, one of the key processes, the formation of bound excitons out of a quasi-free electron-hole plasma \cite{siantidis2001dynamics,kira2004terahertz}, has not yet been theoretically investigated in TMDs. In standard luminescence experiments the semiconductor is excited with frequencies above the single particle band gap. Subsequently, the excited carriers relax via the emission of phonons and emit light corresponding to the energy of the excitonic ground state, which is far below the band gap. Recent experiments \cite{steinleitner2017direct, cha20161s} revealed that the corresponding energy dissipation mechanisms is highly efficient in TMD monolayers, yielding a formation of 1s excitons after off-resonant excitation on a picosecond timescale. The underlying microscopic mechanism enabling this ultrafast relaxation of excitons has not been investigated in literature yet. 

In this work, we provide a fully quantum mechanical description giving access to the momentum-, energy- and time-resolved kinetics of electron-hole correlations. In particular, our model allows to track the phonon-induced relaxation of excited electron-hole pairs into their excitonic ground state after an optical excitation with frequencies far above the 1s resonance. We apply the developed approach to map the relaxation dynamics of excitons in monolayer $\text{MoSe}_2$ on a $\text{SiO}_2$ substrate after an optical excitation $\unit[20]{meV}$ below the single particle band gap. We predict an ultrafast relaxation into the 1s ground state with a relaxation time of about $\unit[1.1]{ps}$. Further, we use the gained insights into the momentum- and energy-resolved exciton dynamics to calculate experimentally accessible ultrafast pump-probe spectra. Here, we reveal that the transient occupation of energetically higher excitonic states leads to the emergence of highly interesting features in the pump-induced response including transient gain signatures and features stemming from dark inter-valley excitons.

\section{Microscopic Model} 

\begin{figure}[!t]
\includegraphics[width=80mm]{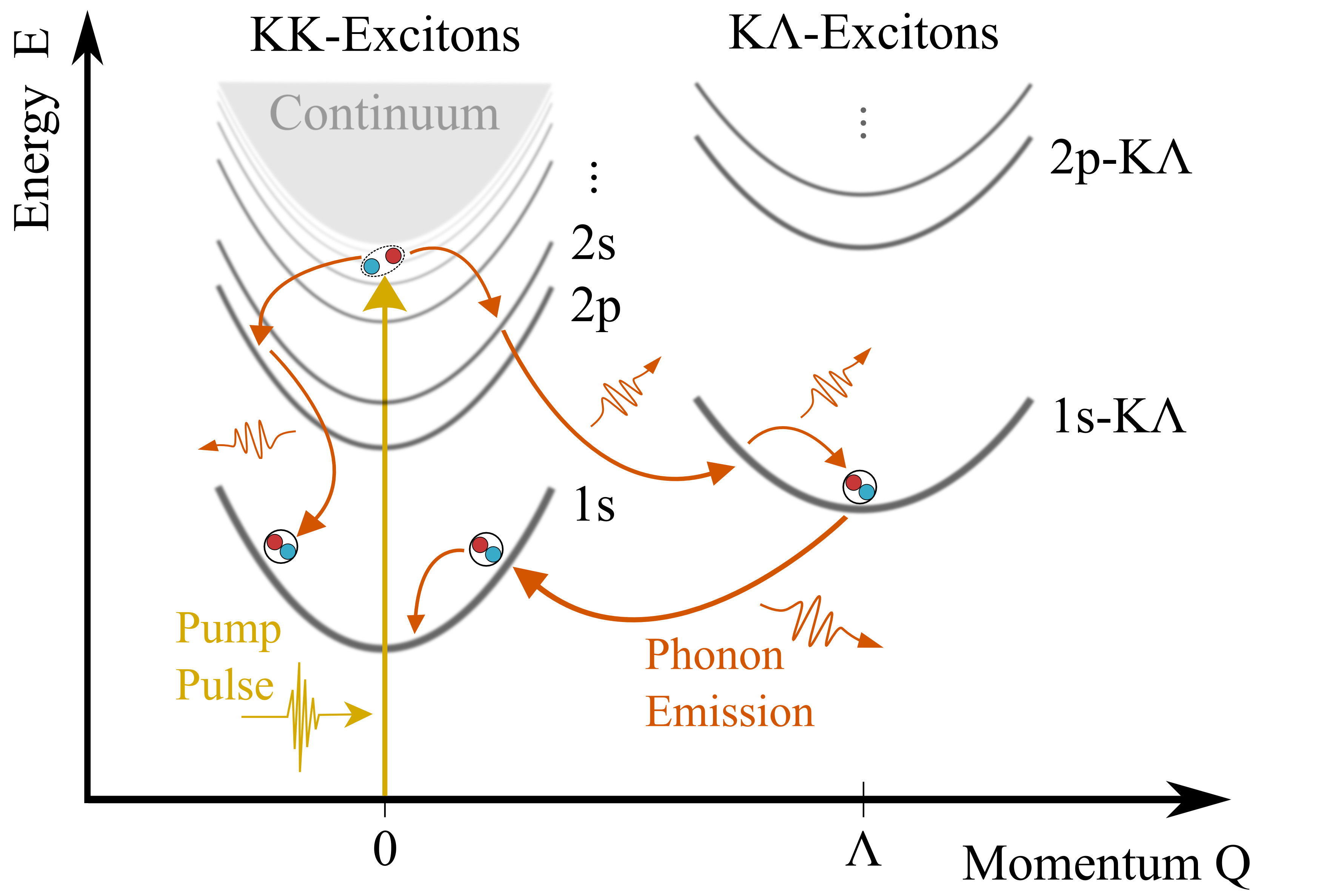} 
\caption{\textbf{Intra-excitonic relaxation cascade in MoSe$_2$.} 
 A pump pulse with an energy close to the single-particle band gap excites quasi-free electron-hole pairs at the scattering continuum (yellow arrow). The hot electron-hole pairs dissipate energy via a sequence of phonon emissions, performing a cascade-like relaxation through energetically lower lying exciton states (orange arrows) including indirect K$\Lambda$ excitons. 
}
\label{fig:principle} 
\end{figure}

To illustrate the formation process of bound excitons after off-resonant excitation, we transfer the single-particle band structure into a two-particle dispersion, which is strongly modified by the Coulomb-attraction between electrons and holes. Figure \ref{fig:principle} schematically illustrates the excitonic band structure and phonon-assisted relaxation processes in molybdenum-based TMD monolayers. Instead of two parabolic bands for electron- and hole momentum, we obtain a Rydberg-type series of states regarding the relative motion of electron and hole, denoted with 1s, 2p, 2s, ..., where each orbital additionally exhibits a quadratic dispersion resulting from the free center-of-mass motion of the exciton. 

Apart from the energetically bound exciton states, there is also a continuum of scattering states with energies above the single-particle band gap, representing quasi-free electron-hole plasma states. Moreover, excitons can be composed of electrons and holes located either at the same valley (KK excitons) or at  different valleys of the Brillouin zone  \cite{Qiu2015,Wu2015,Selig2016,feierabend2017proposal,Selig2017, christiansen2017phonon, malic2017dark}. The indirect intervalley excitons exhibit different effective  masses and binding energies and further show a dispersion minimum at $\mathbf{Q} \neq 0$, which is illustrated in Fig.\ref{fig:principle} for $K\Lambda$ excitons, where the hole is located at K-valley and the electron at the $\Lambda$-valley. Note that excitons with $Q \neq 0$ are dark, i.e. they can neither be excited optically nor decay radiatively, since photons provide only a negligible momentum transfer. In addition to this so called momentum-forbidden dark states, there are also angular-momentum-forbidden (relative motion with angular momentum $l \neq 0$) \cite{berghauser2016optical} and spin-forbidden dark states (electrons and holes with different spin)\cite{Zhang2016,echeverry2016splitting, glazov2014exciton}.

In this work, we model the situation, where a pump pulse with an energy close to the single-particle band gap excites electron-hole pairs at the scattering continuum at $\mathbf{Q}=0$, cf. yellow arrow in Fig.\ref{fig:principle}. These hot electron-hole pairs subsequently dissipate their excess energy via  emission of phonons, performing a cascade-like relaxation through energetically lower lying exciton orbitals, cf. orange arrows in Fig. \ref{fig:principle}. Due to this process, the excited quasi-free electron-hole plasma cools down and forms a population of bound 1s excitons. 
In the following we present a theoretical model, which predicts the above described mechanism based on a well established many-body quantum theory and allows to track the full relaxation kinetics of excitons after off-resonant excitation. We use an equation-of-motion approach to derive the dynamics of a coupled electron, phonon and photon system \cite{Kira2006, rossi2002theory, Kochbuch,Thranhardt2000}. We solve the Heisenberg equation for interband coherences $P_\mathbf{k}= \langle c^{\dagger}_\mathbf{k}  v^{\phantom\dagger}_\mathbf{k}\rangle$ and electron-hole correlations $N_\mathbf{k k' Q} = \delta \langle c^{\dagger}_\mathbf{k+Q}  v^{\phantom\dagger}_\mathbf{k} v^{\dagger}_\mathbf{k'- Q} c^{\phantom\dagger}_\mathbf{k'}\rangle$, where we have introduced operators which create (annihilate) valence band electrons $v^{\dagger}_\mathbf{k}$ ($v^{\phantom\dagger}_\mathbf{k}$) and conduction band electrons $c^{\dagger}_\mathbf{k}$ ($c^{\phantom\dagger}_\mathbf{k}$) with momentum $\mathbf{k}$. To obtain access to the relaxation dynamics of the system after an optical excitation, we take into account electron-phonon, electron-photon and electron-hole interactions. 

In contrast to conventional semiconductors, the effective Coulomb interaction between charge carriers in TMDs leads to a strong coupling of interband coherences at different momenta, yielding an excitonic eigen spectrum for interband transitions. To calculate the energies $E^\mu_0$ and wave functions ${\Phi^\mu_\mathbf{k}}$ of these excitonic eigenmodes $\mu$, we solve the Wannier equation \cite{Kochbuch,Knorr1996,Axt2004,Kira2006,Berghauser2014}
$
 \frac{\hbar^2 k^2}{2 m_{\text{r}}} \Phi^\mu_\mathbf{k} - \sum_\mathbf{q} V_\mathbf{q} {\Phi^\mu_\mathbf{k+q}} = (E^\mu_0-E_{\text{g}})\Phi^\mu_\mathbf{k} .
$
Here, the reduced mass $m_{\text{r}}=(1/m_\text{c}+1/m_\text{v})^{-1}$ is given by the effective masses of valence- ($m_\text{v}$) and conduction band ($m_\text{c}$), while $E_\text{g}$ denotes the single particle band gap \cite{Kormanyos2015}. Moreover, we take into account the dielectric environment of the two-dimensional monolayer by applying a Keldysh potential $V_\mathbf{q}$ for the Coulomb interaction \cite{Ritova1967,Keldysh1978,Berghauser2014}. 
Based on the solutions of the Wannier equation we transform all electronic observables to the excitonic basis, by expanding their momentum dependence in terms of exciton wave functions. Therewith we get access to the excitonic polarisation $P^\mu=\sum_\mathbf{k} {\Phi^\mu_\mathbf{k}}^\ast P_\mathbf{k}$ (coherent excitons) and the exciton occupations $N^\mu_\mathbf{Q} = \sum_\mathbf{k k'} {\Phi^{\mu\ast}_\mathbf{k+\beta Q} }  \Phi^\mu_\mathbf{k'-\alpha Q} N_\mathbf{k k' Q}$ (incoherent excitons) of the state $\mu$ and center-of-mass momentum $\mathbf{Q}$  \cite{Kochbuch,Thranhardt2000,Berghauser2014}. Here, the mass weights $\alpha=m_c/(m_c+m_v)$ and  $\beta=m_v/(m_c+m_v)$ account for the different curvatures in valence and conduction band. In excitonic basis the dynamics of the system is given by
\begin{eqnarray}
\dot{P}^\mu & = & \bigg(\frac{i}{\hbar}E^\mu_0-\gamma_\text{rad}^\mu-\dfrac{1}{2} \sum_{\nu \mathbf{Q}} \Gamma^{\mu \nu}_\mathbf{0Q}\bigg) {P}^\mu +i\Omega^\mu  \label{eq:P_dot},\\
\dot{N}^\mu_\mathbf{Q}  & = &  \sum_{\nu} \Gamma^{\nu \mu}_\mathbf{0Q} \lvert{P}^\nu\rvert^2 -2 \gamma_\text{rad}^\mu \delta_\mathbf{Q,0}  {N}^\mu_\mathbf{Q}  +    \dot{N}^\mu_\mathbf{Q}\rvert_{scat}, \label{eq:N_dot}
\end{eqnarray}
 with $\dot{N}^\mu_\mathbf{Q}\rvert_{scat}= \sum_{\nu \mathbf{Q'}} \Gamma^{\nu \mu}_\mathbf{Q'Q} {N}^\nu_\mathbf{Q'} - \Gamma^{\mu \nu}_\mathbf{QQ'} {N}^\mu_\mathbf{Q}$. The first equation reflects the evolution of coherent excitons, which are initially excited by the optical pump pulse $\mathbf{A}(t)$ determining the excitonic Rabi frequency $\Omega^\mu(t)=ie_0/m_0 \sum_k {\Phi^\mu_\mathbf{k}}^\ast \mathbf{M_k} \cdot \mathbf{A}(t)$. Here, $\mathbf{M_k}$ denotes the optical matrix element for interband transitions \cite{Berghauser2014}, whereas $e_0$ and $m_0$ are the elementary charge and the electron rest mass, respectively. The optically generated polarizations decay radiatively with the rate $\gamma_\text{rad}^\mu$ \cite{Moody2015,Selig2016} as well as via non-radiative dephasing due to the interaction with phonons and form incoherent excitons $N^\mu_\mathbf{Q}$ that are described in the second equation. The probability rate to scatter from the state ($\mu$,$\mathbf{Q}$) to the ($\nu$,$\mathbf{Q'}$) is given by 
\begin{eqnarray}
 \Gamma^{\mu \nu}_\mathbf{QQ'}= \frac{2\pi}{\hbar} \sum_{\pm,\lambda}  \lvert G^{\mu \nu}_{\lambda,\mathbf{Q-Q'}}\rvert^2 \hat{n}^{\lambda, \pm}_\mathbf{Q-Q'}  \delta(\Delta E^{\nu\mu, \lambda}_\mathbf{QQ'}). \label{eq:scattRate} 
\end{eqnarray}
with $\hat{n}^{\lambda, \pm}_\mathbf{Q-Q'}=\frac{1}{2}\pm\frac{1}{2}+n^\lambda_\mathbf{Q-Q'}$ and 
$\Delta E^{\nu\mu, \lambda}_\mathbf{QQ'}=
E^\nu_\mathbf{Q'}-E^\mu_\mathbf{Q} \pm \hbar {\omega}^\lambda_\mathbf{Q-Q'}$. The appearing sum over $+$ and $-$ accounts for phonon emission and absorption, respectively, while ${\omega}^\lambda_\mathbf{q}$ and $n^\lambda_\mathbf{q}$ stand for the phonon frequency \cite{Li2013} and the occupation in the mode  $\lambda$ and momentum $\mathbf{q}$. The delta function accounts for the energy conservation during the scattering processes. We have applied the Markov approximation for carrier-phonon triplets and therefore only allow strictly energy conserving scattering processes \cite{Kochbuch,Thranhardt2000}.
Finally, the exciton-phonon scattering cross section is given by the matrix element  $G^{\mu \nu}_{\lambda,\mathbf{q}}=\sum_\mathbf{k} \Phi^{\mu\ast}_\mathbf{k} \big( \Phi^\nu_\mathbf{k+\beta q} g^c_{\lambda \mathbf{q}} -\Phi^\nu_\mathbf{k-\alpha q} g^v_{\lambda \mathbf{q}}  \big)$,
where $g^c$ and $g^v$ are the electron-phonon matrix elements  in conduction and valence band, respectively \cite{Kaasbjerg2012,Li2013,Jin2014}. 

While the interaction with phonons leads to a decay of the excitonic polarisation $P^\mu$ in  Eq. \eqref{eq:P_dot}, the exact same terms drive the formation of incoherent exciton populations $N^\mu_\mathbf{Q}$, cf. first term in Eq. \eqref{eq:N_dot}\cite{Thranhardt2000,Selig2017}. This reflects the so called polarization-to-population transfer \cite{koch2006semiconductor,Kira2006}, in which coherent excitons generated at $\mathbf{Q}=0$ lose their phase information and obtain a center-of-mass momentum due to the scattering with a phonon. Due to the Boltzmann-like scattering terms in Eq. \eqref{eq:N_dot}, the incoherent excitons in turn redistribute along all exciton states via emission and absorption of phonons until a thermal equilibrium is reached. Additionally, exciton populations with vanishing center-of-mass momenta can recombine radiatively with the rate $\gamma_{\text{rad}}^\mu$.
Note that Pauli blocking terms have been neglected, which is a good approximation in the low density regime.

\begin{figure}[!t]
\begin{centering}
\includegraphics[width=80mm]{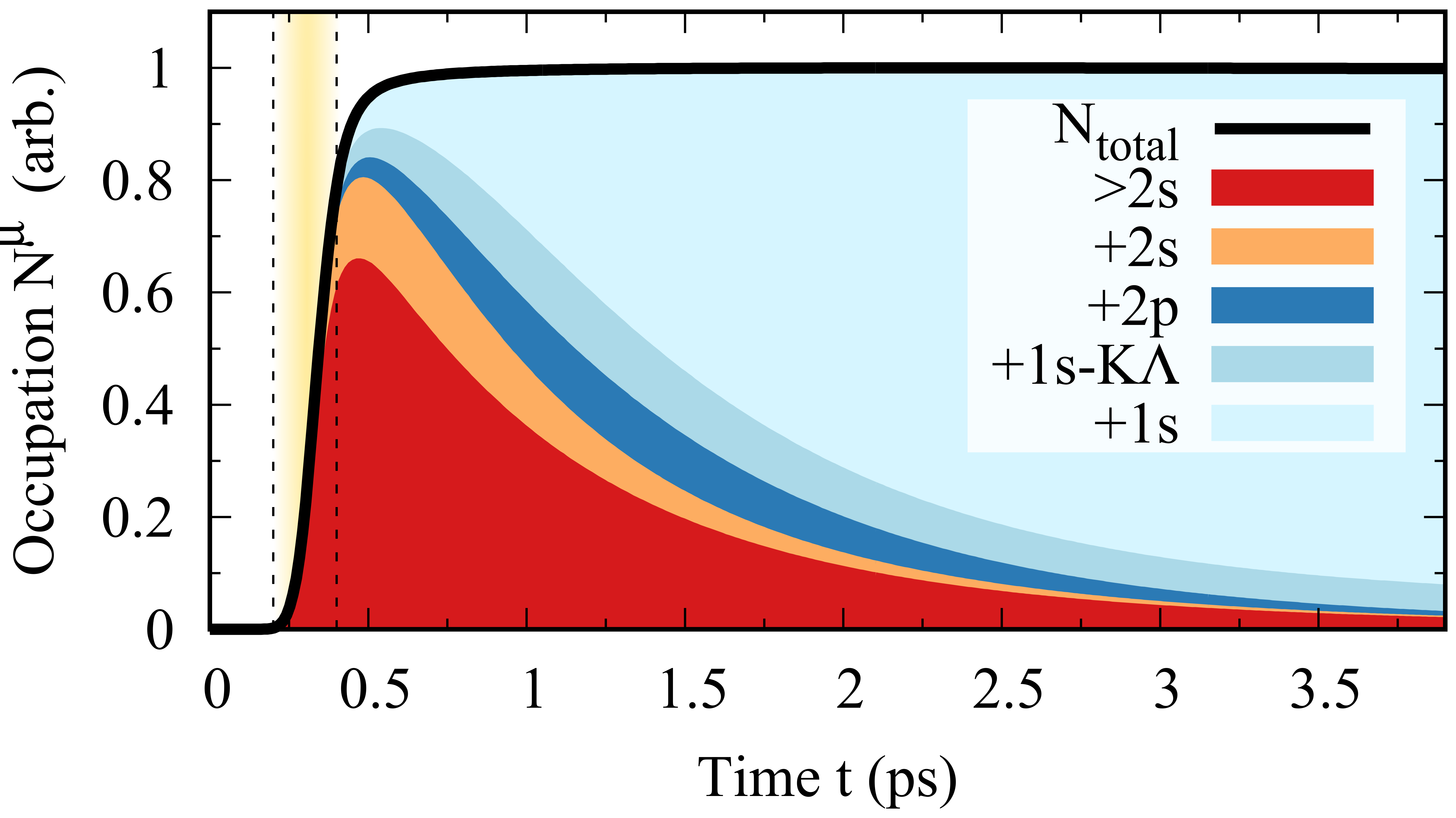} 
\end{centering}
\caption{\textbf{Exciton formation and relaxation.} 
Evolution of the momentum integrated exciton occupations $N^\mu(t)$. While the black line shows the overall number of incoherent excitons, the surface below the black line is divided into colored areas, whose surface ratio represents the relative fraction of the respective exciton state. After the optical excitation close to the single-particle band gap at $t=\unit[300]{fs}$ (yellow area) most electron-hole pairs occupy exciton states above the 2s exciton. Those excited excitons decay into lower lying states via the emission of phonons and at approximately $\unit[1.5]{ps}$ a $1/e$-fraction of the excited pairs has decayed into the 1s-ground state.
}
\label{fig:IntegratedDyn} 
\end{figure}

\section{Exciton Relaxation Dynamics} 

The numerical solution of Eq. \eqref{eq:P_dot} and \eqref{eq:N_dot} provides microscopic access to the time-, energy- and momentum resolved relaxation kinetics of exciton occupations after an optical excitation far above the 1s resonance. In the following, we describe the exciton dynamics in MoSe$_2$ monolayers on $\text{SiO}_2$ at room temperature after an exemplary optical excitation  $\unit[20]{meV}$ below the single particle band gap, which corresponds to the eigen energy of the 5s-state, cf. yellow arrow in Fig. \ref{fig:principle}. Furthermore, we  focus on the occupation dynamics of the so called A-series, i.e. excitons composed of electrons and holes located at the lowest conduction and highest valence band at the K-point, respectively. Moreover, we also include momentum-forbidden dark $K\Lambda$ excitonic states, cf. Fig. \ref{fig:principle}. We take into account the 30 lowest lying direct as well as indirect exciton states  up to the 12s-, 10p- and 9d-state. This allows us to track the exciton dynamics in high resolution.

\begin{figure*}[!t]
\begin{centering}
\includegraphics[width=170mm]{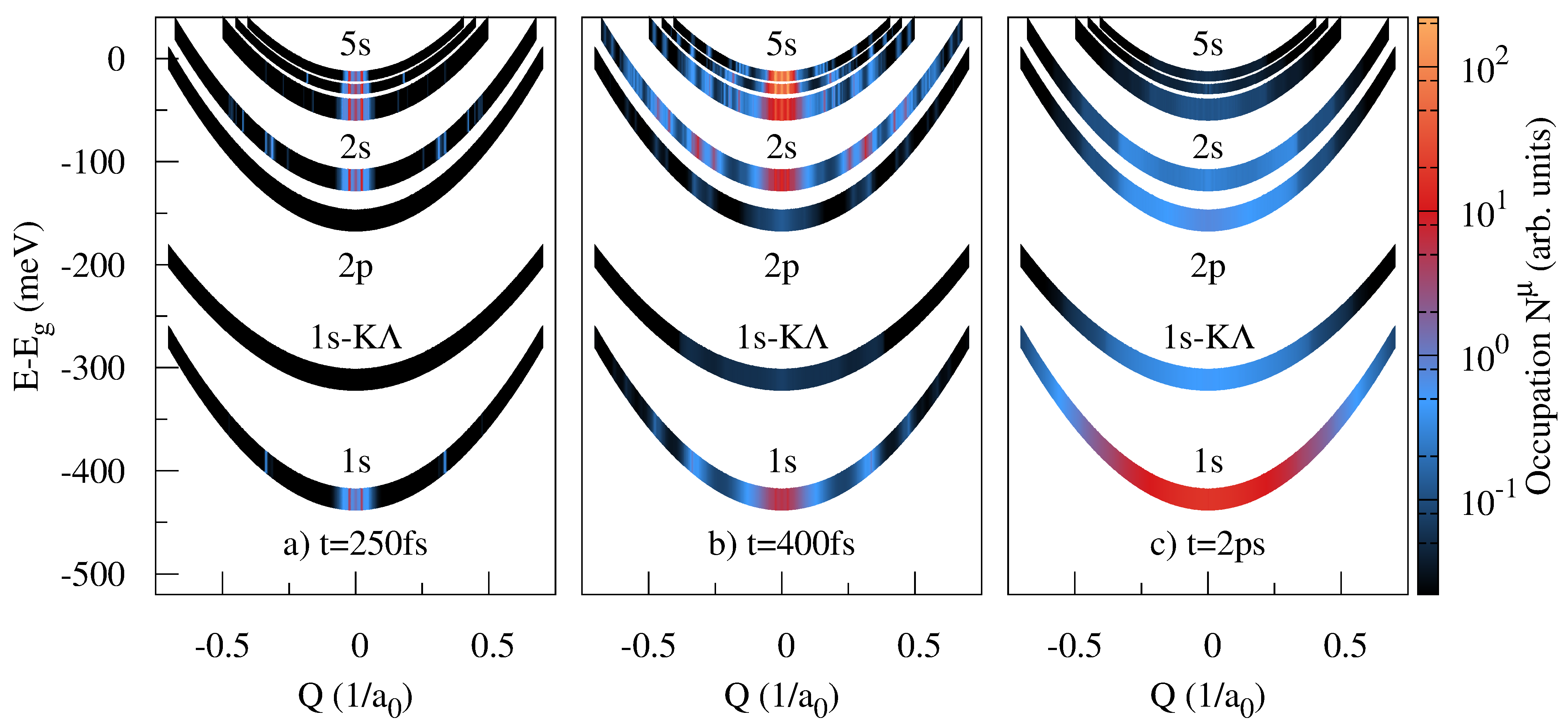} 
\end{centering}
\caption{\textbf{Momentum-resolved exciton dynamics.} 
Momentum- and energy-resolved snapshots of exciton occupations for the most relevant 1s, 1s-K$\Lambda$, 2p, 2s, 3s, 4s and 5s states for three characteristic times. The $\Lambda$-valley has been shifted in momentum for a better presentation. a)  Polarisation-to-population transfer during the pump pulse at $t=\unit[250]{fs}$. b) Intermediate exciton distribution at $t=\unit[400]{fs}$. c) Nearly relaxed Boltzmann-like exciton occupation at $t=\unit[2]{ps}$.
}
\label{fig:MomRes} 
\end{figure*}
 Figure \ref{fig:IntegratedDyn} shows the dynamics of the overall number of excitons (corresponding to the momentum-integrated exciton occupation $N^\mu_\mathbf{Q}$)  together with a color-coded decomposition into the appearing exciton states. Note that the different exciton states were added up successively, so that the colors need to be read similar to a bar chart, representing the fractional contribution of the particular state to the total number of excitons. Further, to obtain a good overview, we have summarized the very dense lying states with energies $E>E_{2s}$.
 The investigated TMD material is excited at $t=\unit[300]{fs}$ by a $\unit[100]{fs}$ long Gauss pulse with a central frequency matching the 5s-eigen energy (yellow shaded region in Fig. \ref{fig:IntegratedDyn}), which leads to the rapid generation of electron-hole pairs (black line). The optical field induces a polarisation, which corresponds to coherent excitons $P^\mu$, cf. Eq. \eqref{eq:P_dot}. The formation of incoherent excitons simultaneously occurs via the so called polarisation-to-population transfer. Here, the interaction with phonons leads to a strong dephasing of the excitonic polarisation $P^\mu$ generating incoherent exciton populations. Hence, the total number of excitons,  increases rapidly during the pump pulse and shortly afterwards, cf. the black line in Fig. \ref{fig:IntegratedDyn}. After about $\unit[1]{ps}$ the remaining polarization has completely decayed into incoherent populations and the total number of excitons stays almost constant due to the large radiative recombination times in the 100-picosecond range.
 The color-coded decomposition into different exciton states shows that during the first few $\unit[100]{fs}$ only states >2s are formed (mainly the resonantly driven 5s- and surrounding states), cf. red fraction in Fig. \ref{fig:IntegratedDyn}. Those higher excitonic states now decay into lower lying states on a picosecond timescale via emission of optical and acoustic phonons. After about $\unit[1.5]{ps}$ a $1/e$-fraction of the total number of excited pairs has decayed into the 1s ground state.

 Since the scattering probability crucially depends on the overlap of excitonic wave functions of the involved states,  scattering between excitons with the same angular momentum is most favorable. The optically excited (bright) s-states preferably decay into the lowest (since in momentum space broadest) s-exciton, given that its dispersion can be reached under emission of a phonon.  An other important aspect, is that the 2s-state very effectively couples to the momentum indirect 1s-$K\Lambda$, which in turn effectively couples back to the direct 1s ground state. As a result, the indirect 1s-$K\Lambda$-state effectively acts as a catalyst, rapidly promoting excitons from the 2s to the 1s-state under emission of two $\Lambda$-phonons, as illustrated in the right relaxation path in Fig \ref{fig:principle}. 

In total, the cascade $ns \rightarrow$ 2s $\rightarrow$ 1s-K$\Lambda$ $\rightarrow$ 1s leads to an ultrafast relaxation of band gap near electron-hole pairs into the excitonic ground state. Thereby, the relaxation cascade mainly occurs via s-states, since the overlaps of wave functions with different angular momentum are small due to phase cancellations. However, due to the large number of densely spaced p- and d-type states, there is also a notable amount of excitons scattered into those states. Due to their small scattering cross sections these anisotropic states slow down the overall relaxation process by trapping excitons in excited states. 

The main features of the described relaxation cascade are illustrated in more detail in Fig. \ref{fig:MomRes}. Here, we present momentum- and energy-resolved snapshots of exciton occupations for the most relevant states 1s, 1s-K$\Lambda$, 2p, 2s, 3s, 4s and 5s. We have chosen three times illustrating   (a) formation, (b) relaxation and (c) thermalization of excitons after a close-to-bandgap excitation.
 Fig. \ref{fig:MomRes}(a) shows the exciton density at $t=\unit[250]{fs}$. Here the pump pulse has just started to induce polarizations, so we can directly observe the optical formation process. In all bright exciton bands, we observe sharp peaks appearing at small momenta, which is characteristic for the polarisation-to-population transfer. Coherent excitons are optically generated within the light cone ($\mathbf{Q} \approx 0$) and are mainly transfered to incoherent populations via absorption of acoustic phonons. The sharp peaks appear at the intersection of phonon dispersion and exciton parabola. The inverse process (emission of acoustic phonons) leads to the accumulation of incoherent excitons at $\mathbf{Q}=0$, which explains the threefold features at small momenta. On the 2s parabola, we also see several stripes at larger momenta, which result from the absorption or emission of an optical phonon. 

Figure \ref{fig:MomRes}(b) shows the exciton distribution at $t=\unit[400]{fs}$ shortly after the pump pulse has past the sample. Here, we still see the pronounced threefold structure at low momenta, but it is now much broader due to interactions with acoustic phonons. Additionally, we see a manifold of peaks at large momenta, resulting from the replication of the sharp peaks at low momenta due to absorption and emission of optical phonons. Especially in the 3s and 2s band we observe a very broad distribution of excitons, which follows from the decay of energetically higher occupations. Notably, the 2p-state is almost empty, since it only couples weakly to the s-states due to the symmetry mismatch. 

Finally, Fig. \ref{fig:MomRes}(c) shows the almost relaxed exciton occupation at $t=\unit[2]{ps}$. The sharp density peaks stemming from the optical formation have been softened due to acoustic phonons and we observe smooth, almost thermalized exciton distributions. While excitons in the 1s ground state are already in thermal equilibrium with the phonon bath giving rise to a Boltzman distribution with a finite occupation of the $K\Lambda$ state, there is still a large amount of excitons trapped in long-living excited states. Especially exciton states with non-vanishing angular momentum are now occupied, which is exemplary shown for the 2p state.


\section{Ultrafast Pump-Probe-Spectroscopy} 

After presenting a microscopic view on the relaxation of excitons in monolayer $\text{MoSe}_2$, we now discuss the possibility to experimentally measure the predicted cascade-like intra-exciton dynamics. Since the investigated processes occur on a picosecond timescale, the best way to capture the dynamics is to use ultrafast sequences of pump and probe pulses. As the investigated quasi-particles are bosons, they do not induce a measurable bleaching of interband absorption via Pauli blocking. However, excitons can absorb/emit light by performing transitions between internal energy levels, as long as they are dipole-allowed, such as transitions from s- to p-states. Those transitions can also occur at center-of-mass momenta larger than zero, since the exciton just travels vertically between the parabolas of two states. Similar to atomic spectroscopy, the absorbed light of a certain frequency is directly proportional to the amount of excitons in initial and final state. The induced optical susceptibility $\Delta \chi$ after the pump pulse is governed by the difference in exciton occupations $N^\nu_\mathbf{Q}, N^\mu_\mathbf{Q}$ of the involved states \cite{Kira2006,koch2006semiconductor}

\begin{eqnarray}
\Delta \chi(\omega) \propto \sum_{\nu \mu \mathbf{Q}} |D^{\nu\mu}|^2 \frac{N^\nu_\mathbf{Q}-N^\mu_\mathbf{Q}}{E^{\mu\nu}-\hbar \omega -i \hbar \gamma^{\nu\mu}_\mathbf{Q}}. \label{eq:response}
\end{eqnarray}
Here $E^{\nu\mu}=E^\nu_0-E^\mu_0$ denotes the transition energy between the exciton $\nu$ and $\mu$, whereas the momentum dependent dephasing is given by the life times of initial and final state $\gamma^{\nu\mu}_\mathbf{Q}=  1/2 \sum_{\alpha \mathbf{Q'}} (\Gamma^{\nu \alpha}_\mathbf{QQ'}+\Gamma^{\mu \alpha}_\mathbf{QQ'})$. Moreover, the oscillator strength of the intra-excitonic transition is given by the square of the dipole matrix element $D^{\nu\mu}$ \cite{Kira2006}.

\begin{figure}[!t]
\begin{centering}
\includegraphics[width=80mm]{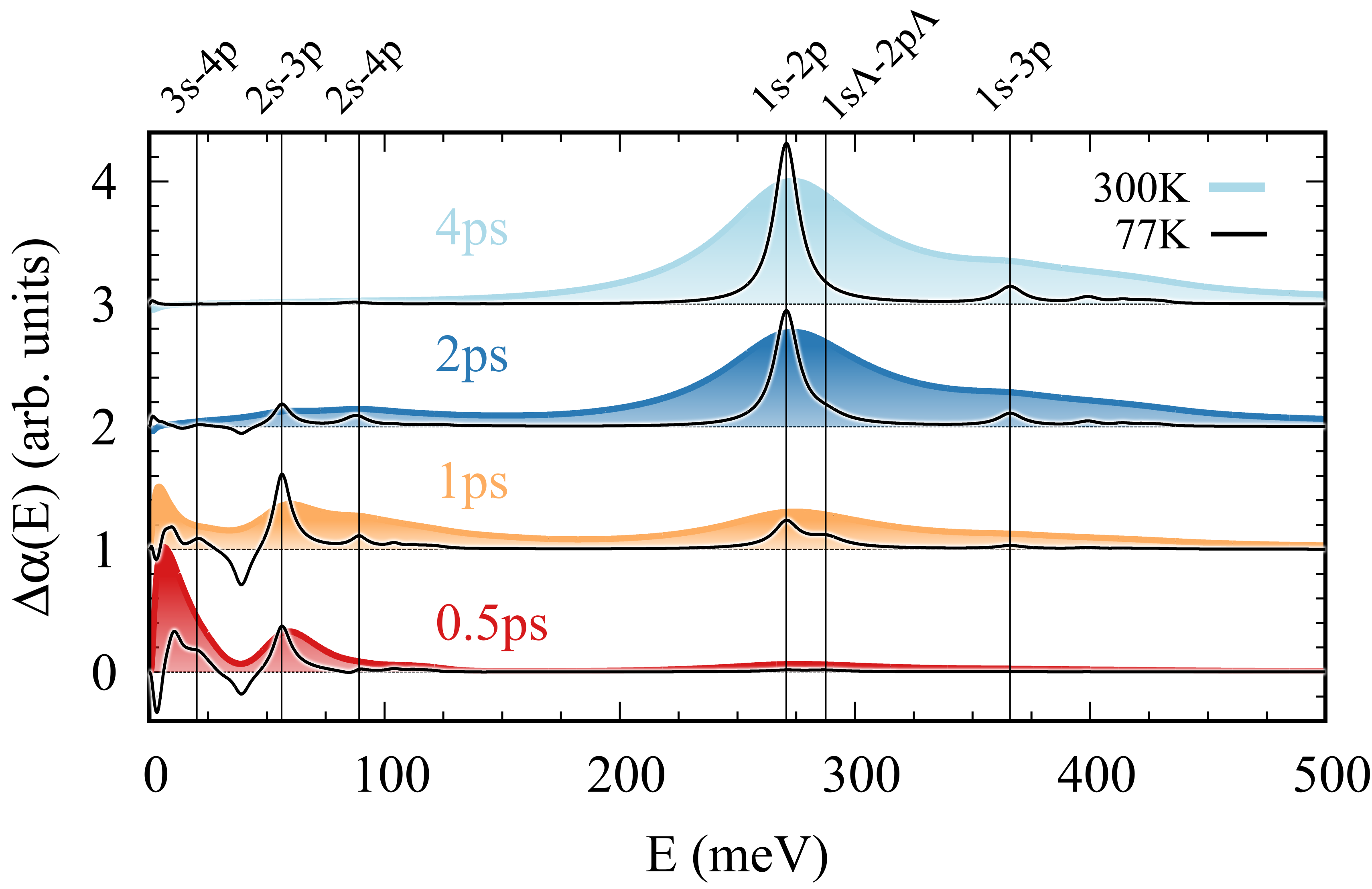} 
\end{centering}
\caption{\textbf{Pump-induced Absorption.} 
Pump-induced absorption as function of the probe energy for four different times. The colored spectra show the calculation at room temperature, whereas the black lines correspond to the same experiment at $\unit[77]{K}$ (scaled by a factor of 0.2). The displayed times correspond to the time axis of the last section.  For early times the occupation of the higher excitonic states yields multiple features in the absorbtion for energies below $\unit[150]{meV}$. At low temperatures, the population inversion between 2s and 2p gives rise to optical gain for energies around $\unit[40]{meV}$. For larger delay times the absorption spectrum is dominated by the response of the 1s-np (n=2,3, etc.) transitions of excitons in the ground state.
}
\label{fig:diffAbsorb} 
\end{figure}

Since the response at a certain frequency directly reflects the involved exciton occupations, the probing of intra-excitonic transition after an interband pump pulse allows to indirectly measure the relaxation dynamics on a femtosecond timescale \cite{kaindl2003ultrafast, pollmann2015resonant, steinleitner2017direct, cha20161s}. To identify experimentally measurable traces of the predicted exciton cascade, we calculate the time-dependent intra-excitonic response function based on the exciton dynamics discussed in the previous section. Figure \ref{fig:diffAbsorb} shows the induced absorption as function of the probe energy for four different times. The colored spectra result from the dynamics presented in the last section (room temperature), while the black curves show the induced absorption at $\unit[77]{K}$.
For short delay times after the pump pulse at $t=\unit[0.5]{ps}$ and $t=\unit[1]{ps}$ (red and orange curve) we observe several absorption peaks for energies below $\unit[150]{meV}$. They predominantly stem from high energy excitons, which occupy s-states above the 2s and perform transitions to energetically close p-states. These spectral features provide direct evidence for the transient but ultrafast occupation of intermediate states during the relaxation from the optically excited close-to-bandgap states. Moreover, we predict that transient population inversions between s- and p-states arise during the relaxation process, since the phonon scattering preferably occurs between states with the same orbital symmetry. For low temperatures, where spectral lines are narrow enough to be clearly distinguished from each other, we observe pronounced gain signals (i.e. negative absorption) e.g. stemming from the 2s-2p transition at about $\unit[40]{meV}$. 
For larger delay times (blue curves in Fig. \ref{fig:diffAbsorb})  we observe a strong increase of the 1s-np resonances stemming from excitons in the ground state. The rise time of this signal provides direct access to the overall relaxation timescale. At low temperatures, we can directly observe the transient occupation of the dark intervalley K$\Lambda$ exciton during the relaxation process.  For short delay times, we predict a clear shoulder on the high energy side of the 1s-2p peak stemming from the corresponding transition in the $\Lambda$ valley, which dissapears after equilibration of the system. At $t=\unit[4]{ps}$ the spectrum is dominated by transitions starting from the 1s-state, reflecting the fully relaxed exciton distribution. In total, Fig. \ref{fig:diffAbsorb} shows that the temporal evolution of the pump-probe spectrum represents a powerful tool to map the ultrafast and cascade-like intra-excitonic relaxation dynamics. Here, upcoming and disappearing peaks can be directly related to a redistribution of excitons along the Rydberg-like series of excitonic states.

\section{Conclusion} 

We have presented a microscopic view on phonon-driven ultrafast relaxation of excited electron-hole pairs into bound excitons. In particular, we have investigated the dynamics of A-excitons in $\text{MoSe}_2$ monolayers on a $\text{SiO}_2$ substrate after an optical excitation $20$ meV below the single particle band gap. Based on the calculated momentum- and energy-resolved exciton dynamics, we have revealed symmetry dependencies within the excitonic relaxation cascade and predict a formation of 1s excitons with a time constant of about $\unit[1.1]{ps}$. Finally, we have presented signatures of intra-exciton dynamics in absorption spectra, which are accessible in  ultrafast pump-probe experiments. For pump-probe delay times smaller than $\unit[2]{ps}$, we predict pronounced multi-peak features for energies below $\unit[150]{meV}$. For low temperatures, we find a clear gain signal resulting from a transient population inversion between 2s and 2p states and a pronounced high energy shoulder of the 1s-2p peak stemming from the intermediate occupation of dark intervalley K$\Lambda$ states. The presented work reveals new microscopic insights into the ultrafast intra-excitonic dynamics in atomically thin 2D materials and provides a concrete recipe for future experimental studies.

\section*{Acknowledgement} 
This project has received funding from the European Union's Horizon 2020 research and innovation programme under grant agreement No 696656. Furthermore, we acknowledge financial support from the Swedish Research Council (VR) and the Deutsche Forschungsgemeinschaft (DFG) through SFB 787 and 951. Finally, we gratefully thank Andreas Knorr (TUB) for fruitful discussions.

\end{document}